\def\Rb{{I\!\! R}}

\def\Cb{\ \hbox{\vrule width 0.6pt height 6pt depth 0pt
              \hskip -3.5 pt} C}

\documentclass{czjphys}         
\usepackage{amsmath}
\usepackage{amssymb}
\usepackage{graphicx}
\begin{document}
\title{Exactly Solvable Many-Body Systems and
Pseudo-Hermitian Point Interactions}
\authori{Shao-Ming Fei\footnote{E-mail: fei@wiener.iam.uni-bonn.de}}

\addressi{Department of Mathematics, Capital Normal University, Beijing
100037\\
Institute of Applied Mathematics, University of Bonn, D-53115
Bonn}

\authorii{}
\addressii{}
\authoriii{}    \addressiii{}
\authoriv{}     \addressiv{}
\authorv{}      \addressv{}
\authorvi{}     \addressvi{}
\headauthor{Shao-Ming Fei} \headtitle{ Exactly Solvable Many-Body
Systems and Pseudo-Hermitian Point Interactions}
\lastevenhead{First Author et al.: Title of the contribution
              ldots}
\refnum{A}
\daterec{XXX}    
\issuenumber{0}
\year{2003}
\setcounter{page}{1}
\firstpage{1}
\maketitle

\begin{abstract}
We study Hamiltonian systems with point interactions
and give a systematic description of the corresponding
boundary conditions and the spectrum properties for
self-adjoint, PT-symmetric systems and systems with real spectra.
The integrability of one dimensional many body systems
with these kinds of point (contact) interactions are investigated
for both bosonic and fermionic statistics.
\end{abstract}

The complex generalization of conventional quantum mechanics has
been investigated extensively in recent years. In particular it is
shown that the standard formulation of quantum mechanics in terms
of Hermitian Hamiltonians is overly restrictive and a consistent
physical theory of quantum mechanics can be built on a complex
Hamiltonian that is not Hermitian but satisfies the less
restrictive and more physical condition of space-time reflection
symmetry (PT symmetry) \cite{bender}. It is proven that if PT
symmetry is not spontaneously broken, the dynamics of a
non-Hermitian Hamiltonian system is still governed by unitary time
evolution. A number of models with PT-symmetric and continuous
interaction potentials have been constructed and studied
\cite{models}. In this article we study Hamiltonian systems with
singular interaction potentials at a point. We give a systematic
and complete description of the boundary conditions and the
spectra properties for self-adjoint, PT-symmetric systems and
systems with real spectra. We then study the integrability of one
dimensional many body systems with these kinds of point interactions.

\section{Self-adjoint point interactions}

Self-adjoint quantum mechanical models describing a particle
moving in a local singular potential concentrated at one or a
discrete number of points have been extensively discussed in the
literature, see e.g. \cite{agh-kh,gaudin,AKbook} and references
therein. One dimensional problems with contact interactions at,
say, the origin ($x=0$) can be characterized by separated or
nonseparated boundary conditions imposed on the wave function
$\psi$ at $x=0$ \cite{kurasov}. The first model of this type
with the pairwise interactions determined by $\delta$-functions
was suggested and investigated in \cite{mcguire}. Intensive
studies of this model applied to statistical mechanics (particles
having boson or fermion statistics) are given in \cite{y,y1}.

Nonseparated boundary conditions correspond
to the cases where the perturbed operator is equal to
the orthogonal sum of two self-adjoint operators in $L_{2}
(-\infty,0]$ and $L_{2} [0,\infty)$.
The family of point interactions for the one dimensional
Schr\"odinger operator $ - \frac{d^2}{dx^2}$
can be described by unitary $ 2 \times  2 $ matrices
via von Neumann formulas for self-adjoint extensions
of symmetric operators.
The non-separated boundary conditions describing the self-adjoint extensions
have the following form
\begin{equation} \label{bound}
\left( \begin{array}{c}
\psi(+0)\\
\psi '(+0)\end{array} \right)
= e^{i\theta} \left(
\begin{array}{cc}
a & b \\
c & d \end{array} \right)
\left( \begin{array}{c}
\psi(-0)\\
\psi '(-0)\end{array} \right),
\end{equation}
where $ad-bc = 1$, $\theta, a,b,c,d \in \Rb$,
$\psi(x)$ is the wave function of a spinless particle with
coordinate $x$. The values $\theta = b=0$, $a=d=1$ in (\ref{bound})
correspond to the case of a positive (resp. negative) $\delta$-function
potential for $c>0$ (resp. $c<0$). For general $a,b,c$ and $d$, the
properties of the corresponding Hamiltonian systems have been studied in
detail, see e.g. \cite{kurasov,ch,abd}.

The separated self-adjoint boundary conditions are described by
\be\label{bounds}
\psi^\prime(+0) = h^+ \psi (+0)~, ~~~
\psi^\prime(-0) = h^- \psi (-0),
\ee
where $h^{\pm} \in \Rb \cup \{ \infty\}$.
$ h^+ = \infty$ or $ h^- = \infty$
correspond to Dirichlet boundary conditions and
$ h^+ = 0$ ~or~ $ h^- = 0$ correspond to
Neumann boundary conditions.

\section{PT-symmetric point interactions}

An operator is said to be PT-symmetric if it commutes with
the product operator of the parity  operator P and the
time reversal operator T. It can be shown that
the family of PT-symmetric second
derivative operators with  point interactions at the origin
coincides with the set of restrictions of the second derivative
operator to the domain of functions
satisfying the boundary conditions at the origin \cite{afkpt}:
\begin{equation} \label{bcond1}
\left(\begin{array}{c}
\psi (+0) \\
\psi' (+0)
\end{array}\right) = B
 \left(\begin{array}{c}
\psi (-0) \\
\psi' (-0)
\end{array}\right)
\end{equation}
for non-separated type, where
$$ B = e^{i \theta}
\left( \begin{array}{cc}
\sqrt{1 +bc} \; e^{i\phi} & b \\
c & \sqrt{1+bc} \; e^{-i\phi}
\end{array} \right),
$$
the real parameters $ b \geq 0, c \geq -1/b$\footnote{If the
parameter $b$ is equal to zero, then the second inequality should be
neglected.}, $ \theta, \phi \in [0, 2 \pi)$; or corresponding to
the separated type
\begin{equation} \label{bcond2}
h_0 \psi' (+0)  =  h_1 e^{i\theta} \psi (+0),~~~ h_0 \psi' (+0) =
 - h_1 e^{-i\theta} \psi (-0)
\end{equation}
with the real phase parameter $ \theta \in [0,2\pi)$ and the
parameter $ {\bf h} = (h_0, h_1)$ taken from the (real) projective
space $ {\bf P}^1$.

The spectrum of any PT-symmetric second
derivative operator with point interactions at the origin
consists of the branch
$[0,\infty)$ of the absolutely continuous spectrum
and at most two (counting multiplicity) eigenvalues, which are
real negative or are (complex) conjugated to each other.
The eigenvalues corresponding to
PT-symmetric eigenfunctions are real and negative.
Every eigenfunction corresponding to any
real eigenvalue can be chosen either PT-symmetric or -antisymmetric.

The spectrum of the PT-symmetric second
derivative operator with non-separated type point interaction at the origin
is pure real if and only if the parameters appearing in
(\ref{bcond1}) satisfy in addition at least one of the following
conditions:
\begin{equation} \label{bcondreal1}
bc \sin^2 \phi  \leq \cos^2 \phi;
\end{equation}
\begin{equation} \label{bcondreal2}
bc \sin^2 \phi  \geq \cos^2 \phi \; \; {\rm and}\; \; \cos \phi
\geq 0.
\end{equation}

\section{Point interactions with real spectra}

We consider further non-separated type boundary
conditions at the origin leading to second derivative operators
with real spectrum. A general form of the boundary condition can be
written as
\begin{equation}\label{generalb}
\left(\begin{array}{c}
\psi (+0) \\
\psi' (+0)
\end{array}\right) = \left( \begin{array}{cc}
\alpha & \beta \\
\gamma & \delta \end{array} \right) \left(\begin{array}{c}
\psi (-0) \\
\psi' (-0)
\end{array}\right),
\end{equation}
where $\alpha$, $\beta$, $\gamma$ and $\delta\in\Cb$.
We suppose that the matrix $ B = \left(
\begin{array}{cc}
\alpha & \beta \\
\gamma & \delta \end{array} \right) $ appearing in the boundary
conditions (\ref{generalb}) is non degenerate (from $GL(2, \Cb)$). Again it is easy
to prove that the operator has branch of absolutely continuous
spectrum $ [0, \infty)$. To study its discrete
spectrum we use the following {\it Ansatz} for the
eigenfunction
$$
\psi (x) = \left\{ \begin{array}{ll} c_1 e^{-ikx}, & x < 0 \\
c_2 e^{ikx}, & x > 0  \end{array} \right. , \; \; \Im k > 0,
$$
corresponding to the energy $\lambda = k^2$. Substituting this
function into the boundary conditions (\ref{generalb}) we get
the dispersion equation $k^2 \beta + ik (\alpha + \delta) - \gamma = 0$.

The set of coefficients $\alpha,~ \beta,~ \gamma$, and $\delta$
satisfying the condition $\Im k_{1,2} \leq 0$
can be parameterized by $8$ real
parameters and leads to operators with pure absolutely
continuous spectrum $ [0, \infty)$. Pure imaginary solutions to
the dispersion equation leads to nontrivial discrete spectrum.
Set $ \tau = \alpha + \delta$.
The solutions are pure imaginary if and only if the
following conditions are satisfied:
\begin{equation}\label{betaneq0}
\tau = t e^{i\theta},~ \beta = b e^{i\theta},~
\gamma = c e^{i \theta},~4 \frac{c}{b} \leq \frac{t^2}{b^2},
\end{equation}
for $\beta\neq 0$, where $ t,b,c $ are real
numbers. If $\beta=0$, the spectrum is guaranteed to be real when
$\alpha + \delta$ and $\gamma$ have the same phases, i.e.,
\begin{equation}\label{beta0}
\tau = t e^{i\theta},~ \gamma = c e^{i\theta}.
\end{equation}

The real spectrum point interaction (\ref{betaneq0})
is parameterized by $6$ real parameters.
The four-parameter family of self-adjoint (non-separated) boundary
conditions (\ref{bound}) is contained in this
$6$-parameter family. The family of PT-symmetric (non-separated)
boundary conditions leading to operators with real spectrum
is also included in the family of boundary conditions (\ref{betaneq0})
or (\ref{beta0}).

\section{Integrable many-body systems}

The self-adjoint boundary conditions (\ref{bound}) and (\ref{bounds}),
PT-symmetric boundary conditions (\ref{bcond1}) and (\ref{bcond2}),
and the real spectrum boundary conditions
(\ref{generalb}) with the parameters satisfying (\ref{betaneq0}) or
(\ref{beta0}) also describe two spinless particles moving in one dimension
with contact interaction when they meet (i.e. the relative coordinate $x=0$).
When the particles have spin $s$ but without any spin coupling among
the particles, $\psi$ represents any one of the
components of the wave function. In the following we study the
integrability of one dimensional systems of $N$-identical particles
with general contact interactions described by the non-separated
boundary conditions that are imposed on the relative coordinates of
the particles. We first consider the case of two particles ($N=2$) with
coordinates $x_1$, $x_2$ and momenta $k_1$, $k_2$ respectively.
Each particle has $n$-`spin' states designated by $s_1$ and $s_2$,
$1\leq s_i\leq n$. For $x_1\neq x_2$, these two particles are free. The
wave functions $\varphi$ are symmetric (resp. antisymmetric) with respect
to the interchange $(x_1,s_1)\leftrightarrow(x_2,s_2)$ for bosons (resp.
fermions). In the region $x_1<x_2$, from the Bethe ansatz the
wave function is of the form,
\be\label{w1}
\varphi=\alpha_{12}e^{i(k_1x_1+k_2x_2)}+\alpha_{21}e^{i(k_2x_1+k_1x_2)},
\ee
where $\alpha_{12}$ and $\alpha_{21}$ are $n^2\times 1$ column matrices.
In the region $x_1>x_2$,
\be\label{w2}
\varphi=(P^{12}\alpha_{12})e^{i(k_1x_2+k_2x_1)}
+(P^{12}\alpha_{21})e^{i(k_2x_2+k_1x_1)},
\ee
where according to the symmetry or antisymmetry conditions,
$P^{12}=p^{12}$ for bosons and $P^{12}=-p^{12}$ for fermions, $p^{12}$
being the operator on the $n^2\times 1$ column that interchanges
$s_1\leftrightarrow s_2$.

Set $k_{12} = (k_1 -k_2)/2$. In the center of mass
coordinate $X=(x_1+x_2)/2$ and the relative coordinate
$x=x_2-x_1$, we get, by substituting (\ref{w1}) and (\ref{w2}) into the
boundary conditions (\ref{generalb}) at $x=0$,
\be\label{a1}
\left\{
\begin{array}{l}
\alpha_{12}+\alpha_{21}
=\alpha P^{12}(\alpha_{12}+\alpha_{21})+
i \beta k_{12}P^{12}(\alpha_{12}-\alpha_{21}),\\
ik_{12}(\alpha_{21}-\alpha_{12})
=
\gamma P^{12}(\alpha_{12}+\alpha_{21})+i\delta k_{12}P^{12}
(\alpha_{12}-\alpha_{21}).
\end{array}\right.
\ee
Eliminating the term $P^{12}\alpha_{12}$ from (\ref{a1})
we obtain the relation
\be\label{2112}
\alpha_{21} = Y_{21}^{12} \alpha_{12}~,
\ee
where
\be\label{a21a12}
Y_{21}^{12}
=\frac{
2ik_{12}(\alpha\delta-\beta\gamma)P^{12}+ik_{12}(\alpha-\delta)+(k_{12})^2\beta+\gamma}
{ik_{12}(\alpha+\delta) + (k_{12})^2\beta-\gamma}.
\ee

For $N\geq 3$ and $x_1<x_2<...<x_N$, the wave function is given by
\be\label{psi}
\ba{rcl}
\varphi&=&\alpha_{12...N}e^{i(k_1x_1+k_2x_2+...+k_Nx_N)}
+\alpha_{21...N}e^{i(k_2x_1+k_1x_2+...+k_Nx_N)}\\
&&+(N!-2)~other~terms.
\ea
\ee
The columns $\alpha$ have $n^N\times 1$ dimensions. The wave functions
in the other regions are determined from (\ref{psi}) by the requirement of
symmetry (for bosons) or antisymmetry (for fermions).
Along any plane $x_i=x_{i+1}$, $i\in 1,2,...,N-1$, from similar
considerations we have
\be\label{a1n}
\alpha_{l_1l_2...l_il_{i+1}...l_N}=Y_{l_{i+1}l_i}^{ii+1}
\alpha_{l_1l_2...l_{i+1}l_i...l_N},
\ee
where
\be\label{y}
Y_{l_{i+1}l_i}^{ii+1}=
\frac{2ik_{l_il_{i+1}}(\alpha\delta-\beta\gamma)P^{ii+1}
+ik_{l_il_{i+1}}(\alpha-\delta) + (k_{l_il_{i+1}})^2 \beta+\gamma}
{ik_{l_il_{i+1}}(\alpha+\delta)+(k_{l_il_{i+1}})^2 \beta-\gamma}.
\ee
Here $k_{l_il_{i+1}}=(k_{l_i}-k_{l_{i+1}})/2$ play the role of spectral
parameters.
$P^{ii+1}=p^{ii+1}$ for bosons and $P^{ii+1}=-p^{ii+1}$ for fermions,
with $p^{ii+1}$ the operator on the $n^N\times 1$ column
that interchanges $s_i\leftrightarrow s_{i+1}$.

For consistency $Y$ must satisfy the Yang-Baxter equation with
spectral parameter \cite{y,ma}, i.e.,
$$
Y^{m,m+1}_{ij}Y^{m+1,m+2}_{kj}Y^{m,m+1}_{ki}
=Y^{m+1,m+2}_{ki}Y^{m,m+1}_{kj}Y^{m+1,m+2}_{ij},
$$
or \be\label{ybe1} Y^{mr}_{ij}Y^{rs}_{kj}Y^{mr}_{ki}
=Y^{rs}_{ki}Y^{mr}_{kj}Y^{rs}_{ij} \ee if $m,r,s$ are all unequal,
and \be\label{ybe2} Y^{mr}_{ij}Y^{mr}_{ji}=1,~~~~~~
Y^{mr}_{ij}Y^{sq}_{kl}=Y^{sq}_{kl}Y^{mr}_{ij} \ee if $m,r,s,q$ are
all unequal. These Yang-Baxter relations are satisfied when
\be\label{realsp}
\alpha\delta-\beta\gamma=1,~~~~~\beta=0,~~~~~\alpha=\delta, \ee
i.e., $\beta=0$, $\alpha=\delta=\pm 1$.

Therefore for self-adjoint contact interactions (\ref{bound}), the
$N$-body system is integrable when $\theta =0$, $a=d=\pm 1$,
$b=0$, $c$ arbitrary. The case $a=d=1$, $\theta =b=0$ corresponds
to the usual $\delta$-function interactions, which has been
investigated in \cite{y,y1}. The case $a=d=-1$, $\theta =b=0$, is
related to a kind of anti-$\delta$ interactions \cite{adf}.

For $N$-body systems with PT-symmetric contact interactions
(\ref{bcond1}), the integrable condition (\ref{realsp}) implies
that $\theta=\phi=b=0$, which is just the usual self-adjoint
$\delta$-interaction.

From (\ref{beta0}) and (\ref{realsp}), we also have that an
many-body system with contact interaction and real spectra is
integrable only when $\alpha=\delta=\pm 1$, $\gamma=c$ for some
$c\in\Rb$, which is again the self-adjoint $\delta$-type
interactions, see Fig.1 for the relations among point interactions
according to integrability.

\begin{figure}[tbp]
\begin{center}
\includegraphics[angle=-90,totalheight=5cm]{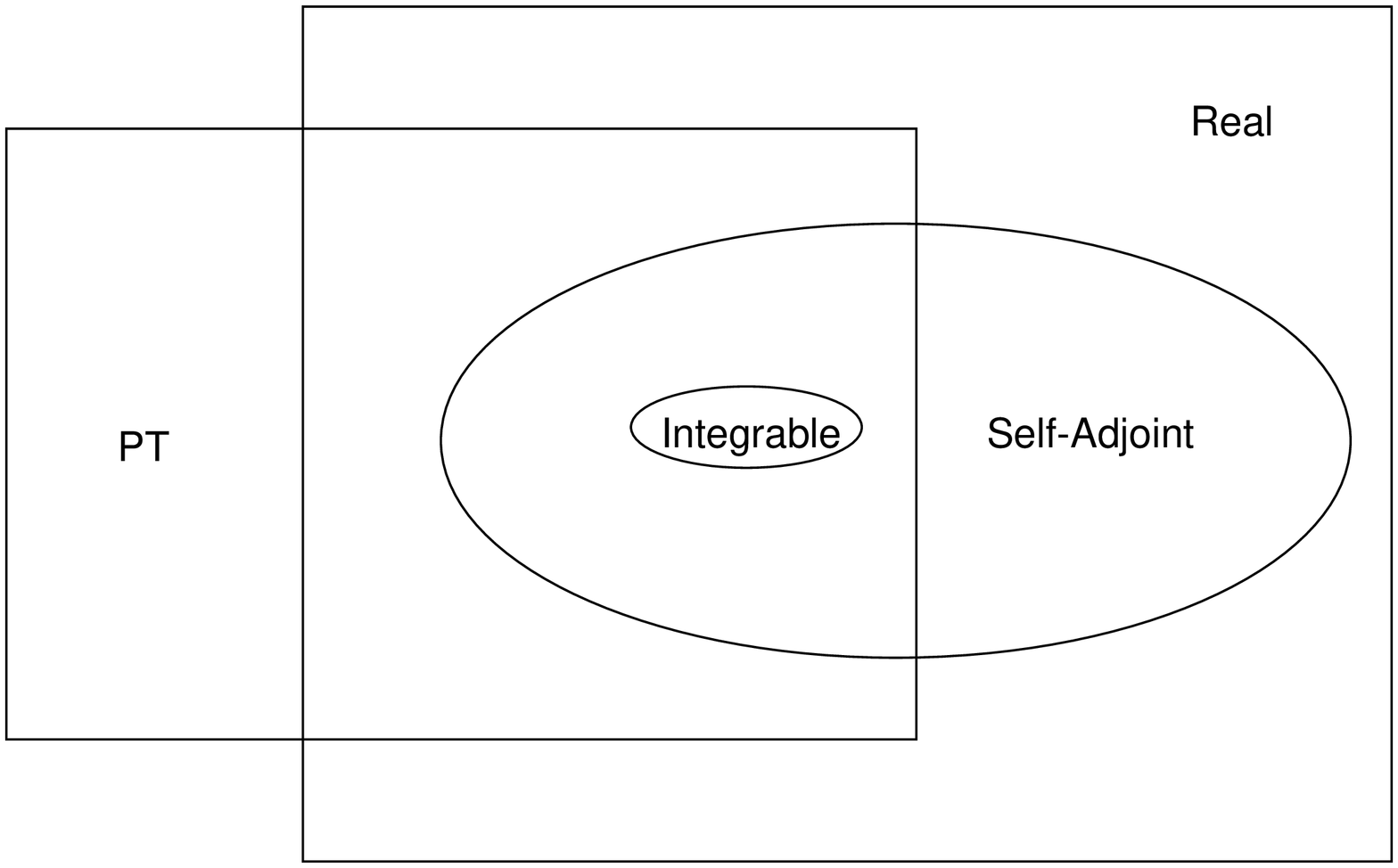}
\end{center}
\caption{Relations among point interactions according to integrability.}
\end{figure}
\bigskip

We have presented a complete picture of self-adjoint,
PT-symmetric and real spectrum point interactions, and
their corresponding integrability. What we concerned here
are just the case of particles with only pure contact interactions,
and the possible contact coupling of the spins of
two particles are not taken into account \cite{spin}.
A further study along this direction would possibly give rise
to more interesting integrable quantum
many-body systems with various symmetries and spectrum properties.


\vspace{2.5ex}
\begin{thebibliography}{20}

\bibitem{bender} C. M. Bender and S. Boettcher, Phys. Rev. Lett. 80,
5243 (1998).\\
C. M. Bender and S. Boettcher and P. N. Meisinger,
J. Math. Phys. 40, 2201 (1999).\\
C. M. Bender, Dorje C. Brody, Hugh F. Jones
Phys. Rev. Lett. 89, 270401 (2002).

\bibitem{models}
F. M. Fern\'{a}ndez, R. Guardiola, J. Ros, and M. Znojil,
J. Phys. A: Math. Gen {\bf 31}, 10105(1998).\\
A. Mostafazadeh, J. Math. Phys. {\bf 43}, 205 (2002);
{\bf 43}, 2814 (2002); {\bf 43}, 3944 (2002).\\
F. Cannata, G. Junker, and J. Trost, Phys. Lett. A {\bf 246}, 219 (1998).\\
E. Delabaere and F. Pham, Phys. Lett A {\bf 250}, 25 and 29 (1998).\\
C. M. Bender, G. V. Dunne, and P. N. Meisenger, Phys. Lett. A {\bf 252}, 272 (1999).\\
C. M. Bender and G. V. Dunne, J. Math. Phys. {\bf 40}, 4616 (1999).\\
M. Znojil, J. Phys. A: Math. Gen {\bf 32}, 7419 (1999).\\
E. Delabaere and D. T. Trinh, J. Phys. A: Math. Gen {\bf 33}, 8771 (2000).\\
A. Khare and B. P. Mandal, Phys. Lett. A {\bf 272}, 53 (2000).\\
B. Bagchi, F. Cannata, and C. Quesne, Phys. Lett. A {\bf 269}, 79(2000).\\
M. Znojil and M. Tater, J. Phys. A: Math. Gen {\bf 34}, 1793 (2001).\\
B. Basu-Mallick and B. P. Mandal, Phys. Lett. A {\bf 284}, 231 (2001).\\
M. Znojil, Phys. Lett. A {\bf 285}, 7 (2001).\\
B. Bagchi, S. Mallik, and C. Quesne, Int. J. Mod. Phys. A {\bf 16}, 2859 (2001).\\
C. M. Bender, G. V. Dunne, P. N. Meisenger, and M. \c{S}im\c{s}ek,
Phys. Lett. A {\bf 281}, 311(2001).\\
Z. Yan and C. R. Handy, J. Phys. A: Math. Gen {\bf 34}, 9907 (2001).\\
Z. Ahmed, Phys. Lett. A {\bf 282}, 343 (2001); {\bf 286}, 231 (2001);
{\bf 294} 287 (2002).\\
C. M. Bender and Q. Wang, J. Phys. A {\bf 34}, 3325 (2001).

\bibitem{agh-kh}
S. Albeverio, F. Gesztesy, R. H\o egh-Krohn and H. Holden, {\it
Solvable Models in Quantum Mechanics}, New York: Springer, 1988.

\bibitem{gaudin}
M. Gaudin, {\it La fonction d'onde de Bethe}, Masson, 1983.

\bibitem{AKbook}
S. Albeverio and R. Kurasov, {\it Singular perturbations of differential
operators and solvable Schr\"odinger type operators},
London Mathematical Society Lecture Note Series, 271,
Cambridge University Press, Cambridge, 2000.

\bibitem{kurasov}
P. Kurasov, J. Math. Analy. Appl.
{\bf 201}, 297-323 (1996).

\bibitem{mcguire} J.B. McGuire,
J. Math. Phys. {\bf 5}, 622-636 (1964); {\bf 6}, 432-439 (1965);
{\bf 7}, 123-132 (1966).\\
J.B. McGuire and C.A. Hurst,
J. Math. Phys. {\bf 13}, 1595-1607 (1972); {\bf 29}, 155-168 (1988).

\bibitem{y}
C.N. Yang, Phys. Rev. Lett. {\bf 19}, 1312-1315 (1967).\\
C.N. Yang, Phys. Rev. {\bf 168}, 1920-1923 (1968).

\bibitem{y1}
C.H. Gu and C.N. Yang, Commun. Math. Phys. {\bf 122}, 105-116
(1989).

\bibitem{ch}
P. Chernoff and R. Hughes, J. Func. Anal. {\bf 111}, 97-117 (1993).

\bibitem{afkpt}
S. Albeverio, S.M. Fei and P. Kurasov,
Lett. Math. Phys {\bf 59}, 227-242 (2002).

\bibitem{abd}
S. Albeverio, Z. Brze\'{z}niak and L D\c{a}browski,
J. Phys. A{\bf 27}, 4933-4943 (1994).

\bibitem{adf}
S. Albeverio, L. D{\c a}browski and S.M. Fei,
Int. J. Mod. Phys. B {\bf 14}, 721-727 (2000).

\bibitem{ma}
Z.Q. Ma, {\it Yang-Baxter Equation and Quantum Enveloping Algebras},
World Scientific, 1993.\\
V. Chari and A. Pressley, {\it A Guide to Quantum Groups}, Cambridge
University Press, 1994.\\
C. Kassel, {\it Quantum Groups}, Springer-Verlag, New-York, 1995.\\
S. Majid, {\it Foundations of Quantum Group Theory}, Cambridge
University Press, 1995.\\
K. Schm\"udgen, {\it Quantum Groups and Their Representations},
Springer, 1997.

\bibitem{spin}
S. Albeverio, S.M. Fei and P. Kurasov,
Rep. Math. Phys. {\bf 47}, 157-165 (2001).

\end{thebibliography}
\end{document}